\documentclass[twocolumn,showpacs,preprintnumbers,amsmath,amssymb]{revtex4}
\usepackage{graphicx}% Include figure files
\usepackage{dcolumn}% Align table columns on decimal point
\usepackage{bm}% bold math

\begin{document}

\preprint{APS/123-QED}

\title{Effect of energy deposited by cosmic-ray particles on 
interferometric gravitational wave detectors}% Force line breaks with \\

\author{Kazuhiro Yamamoto}
 \email{kazuhiro.yamamoto@aei.mpg.de}
 \altaffiliation[Present address: ]{Max Planck Institute 
for Gravitational Physics,
Albert Einstein Institute,
Callinstrasse 38, D-30167 Hannover, Germany.}
%Lines break automatically or can be forced with \\
\author{Hideaki Hayakawa}
\author{Atsushi Okada}
\author{Takashi Uchiyama}
\author{Shinji Miyoki}
\author{Masatake Ohashi}
\author{Kazuaki Kuroda}
\affiliation{%
Institute for Cosmic Ray Research, the University of Tokyo,
5-1-5 Kashiwa-no-Ha, Kashiwa, Chiba 277-8582, Japan}%

\author{Nobuyuki Kanda}
\affiliation{%
Department of Physics, Osaka City University, 
3-3-138 Sugimoto, Sumiyoshi-ku, Osaka, Osaka 558-8585, Japan}%

\author{Daisuke Tatsumi}
\author{Yoshiki Tsunesada}
\affiliation{%
National Astronomical Observatory of Japan,
2-21-1 Osawa, Mitaka, Tokyo 181-8588, Japan}%

\date{\today}% It is always \today, today,
             %  but any date may be explicitly specified

\begin{abstract}
We investigated the noise 
of interferometric gravitational wave detectors 
due to heat energy deposited by cosmic-ray particles. 
We derived a general formula that describes the response 
of a mirror against a cosmic-ray passage.
We found that there are differences in the comic-ray responses 
(the dependence of temperature and cosmic-ray track position) 
in cases of interferometric and resonant gravitational wave detectors. 
The power spectral density of vibrations caused by 
low-energy secondary muons 
is 100-times smaller than the 
goal sensitivity of future second-generation interferometer projects, such as 
LCGT and Advanced LIGO. The arrival frequency of high-energy cosmic-ray muons 
that generate enough large showers inside mirrors of LCGT and Advanced LIGO 
is one per a millennium. 
We also discuss the probability of exotic-particle detection with 
interferometers.  
\end{abstract}

%interferometric gravitational wave detector, cosmic ray, muon, 
%exotic-particle search

%95.55.Ym,14.80.-j, 
% PACS, the Physics and Astronomy
% Classification Scheme.
%\keywords{Gravitational wave, Interferometric detector, 
%Cryogenic detector, Frequency stabilization, 
%Thermal noise, Mirror, 
%Mechanical loss, Dielectric multilayer reflective coating, 
%Ion-beam sputtering, SiO$_2$/Ta$_2$O$_5$}
%Use showkeys class option if keyword
%display desired

\maketitle

\section{Introduction}

Recent improvements of the sensitivity and operational stability
of gravitational wave detectors is remarkable.
Observation runs 
have already been performed in several 
interferometer (LIGO \cite{LIGO}, VIRGO \cite{VIRGO}, 
GEO \cite{GEO}, TAMA \cite{TAMA}, CLIO \cite{CLIO}) and 
resonator (ALLEGRO \cite{ALLEGRO}, EXPLORER \cite{EXPLORER}, 
NAUTILUS \cite{NAUTILUS}, AURIGA \cite{AURIGA}, NIOBE \cite{NIOBE}, 
MARIO SCHENBERG \cite{SCHENBERG}) 
projects. In order of gravitational wave detection, 
the reduction of noise and fake triggers is crucial, since the gravitational 
wave amplitude and number of events are expected to be small and rare. 
In 1969, it was pointed out that cosmic-ray particles could cause 
fake triggers in resonators \cite{Beron}. 
The interpretation of excitations of resonators 
by cosmic-ray particles is follows: 
The heat energy deposited by cosmic-ray particle passages induces 
temperature gradients around their tracks, and  
thermal stress excites internal vibrations of the resonator. 
These phenomena have been investigated, for example,  
observations of excited 
resonator vibrations by beams from accelerators \cite{Strini,Albada}, 
simultaneous detection of resonator excitation and cosmic-ray particles 
\cite{Astone1}, 
and studies of exotic events in super-conductive resonators 
\cite{Astone2,Astone3,RAP}. 
In some research, resonators were also operated and treated as exotic-particle 
detectors \cite{Bernard,Liu,Astonenuclearite,Foot}.
These studies suggest that cosmic-ray heating is a possible noise source
in interferometric detectors 
\cite{Giazotto,Clay,Braginsky-cosmic,Braginsky-cosmic2}
(other effects on interferometers, 
the momentum and electrical charge brought by cosmic-ray particles, 
are discussed in Refs. \cite{Giazotto,Clay,Braginsky-cosmic, Marin}). 
%The momentum transfer is not a serious problem).

We investigated details of this effect 
by cosmic-ray energy deposition in interferometers.
A formula that describes the response 
of a mirror against a cosmic-ray passage 
was derived. This formula reveals 
differences between the cosmic-ray responses 
of interferometers and resonators. 
We used it to evaluate the amplitude of vibrations caused by 
cosmic-ray particles in typical cases of interferometers, and examined the 
effect in gravitational wave detection.
We also considered the probability of exotic-particle detection with 
interferometers.  

\section{Formula of excited motion by a cosmic-ray particle}

%\subsection{Comparison with resonator}
%The difference between the resonator and interferometer
%(relaxation time)

\subsection{Outline of derivation of the formula}

In order to simplify the discussion, 
the mirror vibration excited by a cosmic-ray particle 
is investigated. A vibration excited by many particles, like 
a shower, is a superposition of that by one particle. 
The excitation by a particle is considered under the following assumptions.  
The particle goes straight and never stops in the mirror.  
Its speed is faster than that of sound in the mirror.
A long and narrow heated volume appears 
at the instant of particle passage. 

The heat-conduction equation is solved in order to calculate the 
time evolution of the temperature gradient. 
The vibration of the mirror is examined using the equation of 
motion of an elastic body with thermal stress, 
which is proportional to the thermal gradient. 

\subsection{Formula}

%\subsubsection{Temperature}

Since the heated volume is smaller than that of the mirror, 
itself 
%(the order of ten centimeter) 
\cite{initial1,initial2}, 
%({\bf really?}), 
the mirror and the initial heated volume are treated 
as an infinite body and a line, respectively. 
The direction of the cosmic-ray track is taken as the $z$-axis. 
%The initial value of the function, $\delta T(r,\theta,z,t)$, 
%which is the temperature difference caused by a cosmic-ray particle  
%is assumed as 
%\begin{equation}
%\delta T (r,\theta,z,0) = \delta T_0 \exp \left(-\frac{r^2}{b^2}\right).
%\label{initial T}
%\end{equation}
%The quantitiy, $b$, is the radius of the heated volume. 
The heat-conduction equation is described as \cite{Landau} 
\begin{equation}
\frac{\partial }{\partial t} \delta T -\frac{\kappa}{\rho C} \Delta (\delta T)
= \frac1{\rho C} \frac{dE}{dl} \delta(x) \delta(y) \delta(t),
\label{conductivity}
\end{equation}
where $\delta T$ 
is the temperature difference caused by a cosmic-ray particle. 
The quantities $\kappa, \rho, C$ and $dE/dl$ are the thermal conductivity, 
density, specific heat per unit mass, and 
energy loss of a particle per unit length, respectively. 
The solution is described as \cite{thermal solution}
%\begin{equation}
%\delta T(r,\theta,z,t) = \frac{\delta T_0}{t/\tau_b+1}
%\exp\left(-\frac{r^2}{b^2}\frac1{t/\tau_b+1}\right),
%\label{delta T}
%\end{equation}
%\begin{equation}
%\tau_b = \frac{\rho C b^2}{4 \kappa}\label{tau}.
%\end{equation}
\begin{equation}
\delta T = \frac{1}{4\pi \kappa t}
\frac{dE}{dl}
\exp\left[-\frac{\rho C}{4 \kappa t}\left(x^2+y^2\right)\right].
\label{delta T}
\end{equation}
%The examples of this solution are in Fig. \ref{deltaT}. 
The radius of the heated volume increases with time due to conduction. 
The time when the heated area radius becomes $a$ is 
\begin{equation}
\tau_a = \frac{\rho C a^2}{4 \kappa}.
\label{ta}
\end{equation}
%\begin{figure}
%\includegraphics[width=8.6cm]{deltaT}
%\caption{\label{deltaT}The diffusion of the heat energy deposited 
%by a cosmic-ray particle, 
%Eq. (\ref{delta T}). 
%The quantitiy, $b$, is the radius of the heated volume 
%at the initial time ($t=0$). 
%The temperature ($\delta T$) at the cosmic-ray track ($r=0$) 
%and the heated radius become lower and larger, respectively.}
%\end{figure}

%\subsubsection{Elastic vibration}

The equation of motion of an elastic body with thermal stress 
is described as \cite{Landau}
\begin{eqnarray}
\rho \frac{\partial^2 \boldsymbol{u}}{\partial t^2}
- \frac{Y}{2(1+\sigma)}\Delta \boldsymbol{u} 
&-& \frac{Y}{2(1+\sigma)(1-2\sigma)} {\rm grad\  div} \boldsymbol{u}
\nonumber\\ 
&=& -\frac{Y \alpha}{1-2\sigma} {\rm div} \delta T,
\label{eq of motion} 
\end{eqnarray}
where $\boldsymbol{u}$ represents the displacement of a volume 
element in the elastic body.
The quantities $Y,\sigma$ and $\alpha$ are Young's modulus, 
the Poisson ratio, and 
the linear thermal-expansion coefficient, respectively. 
By substituting Eq. (\ref{delta T}) for 
Eq. (\ref{eq of motion}), we obtain the output of a interferometer $X$, 
\begin{equation}
X = \int_{\rm surface} u_{\rm opt}(\boldsymbol{r}) P(\boldsymbol{r}) dS,
\end{equation}
where $u_{\rm opt}$ is the optical axis component of $\boldsymbol{u}$ 
and $P$ is the intensity profile of the laser beam, 
\begin{equation}
P(\boldsymbol{r}) 
= \frac{2}{\pi {r_0}^2}\exp\left(-\frac{2 r^2}{{r_{0}}^2}\right).
\end{equation}
The quantities $r$ and $r_0$ are the distance from the optical axis 
and the beam radius. We employ the modal expansion method  
\cite{Saulson, Yamamoto-D, Yamamoto-mode} 
to calculate $\boldsymbol{u}$ and $X$. 
In this method, $\boldsymbol{u}$ and $X$ are represented by a superposition 
of the resonant mode displacement, 
\begin{eqnarray}
\boldsymbol{u}(\boldsymbol{r},t)
&=& \sum_{n}\boldsymbol{w}_n(\boldsymbol{r})q_n(t),
\label{mode decomposition}\\
X(t) &=& \sum_{n} q_n(t), \label{observed coordinate decomposition}
\end{eqnarray} 
where $\boldsymbol{w}_n$ and $q_n$
represent the displacement and time development of the $n$-th 
resonant mode, respectively.
These basis functions are normalized to satisfy a condition 
\cite{Yamamoto-D,Yamamoto-mode}, 
\begin{equation}
\int_{\rm surface} w_{n,{\rm opt}}(\boldsymbol{r}) 
P(\boldsymbol{r}) dS = 1,
\label{normalized condition}
\end{equation} 
where $w_{n,{\rm opt}}$ is the optical axis component of $\boldsymbol{w}_n$.
The equation of motion of each mode is the same as that 
of a harmonic oscillator, 
\begin{equation}
-m_n \omega^2 \tilde{q}_n(\omega) + m_n {\omega_n}^2 [1+{\rm i}\phi_n(\omega)] 
\tilde{q}_n(\omega)=\tilde{F}_n(\omega),
\label{traditional1}
\end{equation}
in the frequency domain. 
The quantity $\phi_n$ is the loss angle, which represents 
dissipation of the $n$-th mode \cite{Saulson}. 
The force $F_n$ applied on the $n$-th mode is related to the thermal stress. 
The quantities $m_n$ and $\omega_n$ are the effective mass and 
the resonant angular frequency \cite{Yamamoto-D,Yamamoto-mode}.  
The effective mass is defined as 
\begin{equation}
m_n = \int_{\rm volume} \rho \boldsymbol{w}_n(\boldsymbol{r}) \cdot 
\boldsymbol{w}_n(\boldsymbol{r}) dV.
\label{effective mass} 
\end{equation}
The quantities $\tilde{q}_n(\omega)$ and $ \tilde{F}_n(\omega)$ 
are the Fourier components of $q_n$ and $F_n$, respectively: 
\begin{eqnarray}
\tilde{X}(\omega) &=& \frac1{2\pi}
\int_{-\infty}^\infty X(t)\exp(-{\rm i} \omega t) dt, \\
X(t) &=& \int_{-\infty}^\infty \tilde{X}(\omega)\exp({\rm i} \omega t) d\omega.
\end{eqnarray}

The force $F_n$ is obtained from the modal decomposition of 
the thermal stress on the right-hand side of 
Eq. (\ref{eq of motion}).
The decomposition procedure \cite{Yamamoto-D,Yamamoto-mode} 
is as follows. 
The thermal-stress term 
is multiplied by $\boldsymbol{w}_n$. The integral of this 
inner product over all the volume is $F_n$.
%While $a$ is smaller than the wavelength of the mode, 
%this force is independent of the time because the decay of the 
%thermal stress cancels the expansion of the volume 
%where the thermal stress is applied. 
%After the heated volume radius 
%becomes larger than that of the mode wavelength,
%$F_n$ decreases.
This force $F_n$ decreases
after the heated volume scale, $a$, becomes larger than  
the $n$-th mode wavelength. In order to simplify the discussion,
it is assumed that 
the time evolution of $F_n$ \cite{Bernard,Marinho} is expressed as 
\begin{equation}
F_n(t) = \left\{
\begin{array}{cc}
F_n(0) \exp \left(-\frac{t}{\tau_n}\right) & (t > 0) \\
0 & (t<0)
\end{array}\right. .
\label{F_n}
\end{equation}
The quantities $F_n(0)$ and $\tau_n$ are the initial value 
and the decay time of the force, respectively. 
The initial value, $F_n(0)$, is written as \cite{Liu,Liucomment}
\begin{equation}
F_n(0) = \frac{Y\alpha}{1-2\sigma}\frac1{\rho C}
\left(\int {\rm div} \boldsymbol{w}_n dl \right)
\frac{dE}{dl}.
\label{stress-amplitude}
\end{equation}
The integral along the cosmic-ray track represents the coupling between the 
thermal stress and the $n$-th mode. The coefficient $1/(\rho C)$ is 
a factor used to transform the heat energy into the temperature gradient. 
The force $F_n$ is described as a product of the temperature gradient 
and $Y\alpha/(1-2\sigma)$. 
From Eq. (\ref{ta}), the time $\tau_n$ 
when the heated volume radius becomes comparable 
to the wavelength of the $n$-th mode is 
expressed as
\begin{equation}
\tau_n = \frac{\rho C {\lambda_n}^2}{4 \kappa} 
= \frac{\pi^2 \rho C v^2}{\kappa {\omega_n}^2}
\sim \frac{\pi^2 Y C}{\kappa {\omega_n}^2}.
\label{tau n}
\end{equation}
The quantities $\lambda_n$ and $v$ are the wavelength and sound velocity: 
\begin{equation}
\lambda_n = \frac{2 \pi v}{\omega_n}, 
\end{equation}
\begin{equation}
v \sim \sqrt{\frac{Y}{\rho}}.
\label{sound velocity}
\end{equation}
The Fourier component of $F_n$ in Eq. (\ref{F_n}) is written in the form 
\begin{equation}
\tilde{F}_n(\omega) 
= \frac{F_n(0)}{2\pi}\frac{\tau_n}{1+{\rm i}\omega\tau_n}.
\label{tilde F_n}
\end{equation}

We now write down the formula of the mirror vibration excited 
by a cosmic-ray particle using
Eqs. (\ref{observed coordinate decomposition}), (\ref{traditional1}), 
(\ref{stress-amplitude}) and (\ref{tilde F_n}):
\begin{eqnarray}
\tilde{X}(\omega) &=& \sum_n \tilde{q}_n(\omega)\nonumber\\ 
&=& \sum_n \frac{\tilde{F}_n(\omega)}
{-m_n\omega^2+m_n{\omega_n}^2(1+{\rm i}\phi_n)}\nonumber\\
%&=& \frac1{2 \pi}
%\sum_n \frac{F_n(0)}{-m_n\omega^2+m_n{\omega_n}^2(1+{\rm i}\phi_n)}
%\frac{\tau_n}{1-{\rm i}\omega\tau_n}\nonumber\\
&=& \frac1{2 \pi}\frac{Y\alpha}{1-2\sigma}\frac1{\rho C}
\frac{dE}{dl}\nonumber\\
&&\times \sum_n \frac1{-m_n\omega^2+m_n{\omega_n}^2(1+{\rm i}\phi_n)}
\frac{\tau_n}{1+{\rm i}\omega\tau_n}\nonumber\\
&&\times\left(\int {\rm div} \boldsymbol{w}_n dl \right).
\label{tilde X}
\end{eqnarray}

\subsection{Frequency dependence of the formula}

\begin{figure}
\includegraphics[width=8.6cm]{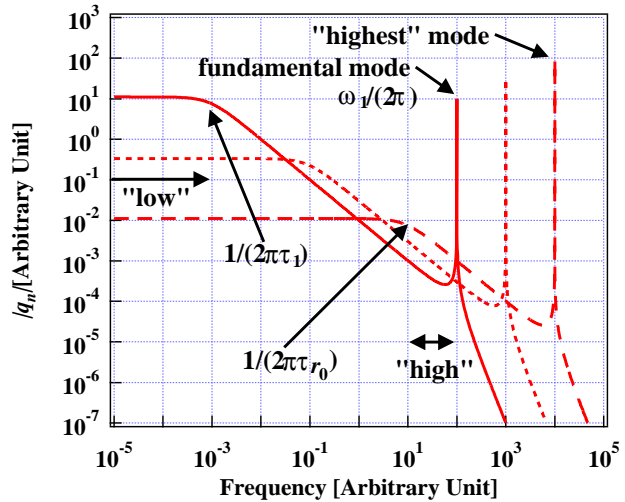}
\caption{\label{Xfreq}A schematic view of the frequency dependence of 
the Fourier components of mode motion excited by a cosmic-ray particle, 
$\tilde{q}_n(\omega)$ in Eq. (\ref{tilde X}).
The absolute value 
$|\tilde{q}_n(\omega)|$ is 
inversely proportional to the frequency between  
$1/(2 \pi \tau_n)$ and $\omega_n/(2 \pi)$. 
Below the cut-off frequency, $1/(2 \pi \tau_n)$, 
it is constant. The "highest" mode is that with 
a wavelength comparable to the beam radius, $r_{0}$, and 
contributions of higher modes 
are negligible in the summation of Eq. (\ref{tilde X})
\cite{Gillespie,Bondu-mode}. The cut-off frequency 
of the "highest" mode, $1/(2 \pi \tau_{r_{0}})$, 
is smaller than the fundamental mode resonant frequency, 
$\omega_1/(2 \pi)$, in general. 
The cut-off frequency, $1/(2\pi\tau_n)$, of the lower mode is smaller.
In the range between $1/(2 \pi \tau_{r_{0}})$ and 
$\omega_1/(2 \pi)$ (the "high" frequency region in this graph),  
$|\tilde{X}(\omega)|=|\sum \tilde{q}_n(\omega)|$ is inversely proportional 
to the frequency. 
Below the cut-off frequency of the fundamental mode, 
$1/(2 \pi \tau_1)$ (the "low" frequency region in this graph), 
$|\tilde{X}(\omega)|$ 
is independent of the frequency. The approximation formulae 
in the "high" and "low" frequency regions, 
Eqs. (\ref{tilde X approx 1}) and (\ref{tilde X approx 2}),  
are derived using Eq. (\ref{tilde X}).}
\end{figure}
A schematic view of the frequency dependence of the modes 
$\tilde{q}_n(\omega)$ 
in Eq. (\ref{tilde X}) is shown in 
Fig. \ref{Xfreq}. 
Here, we discuss the frequency dependence below the 
resonant frequencies of the mirrors, 
because the observation band of interferometers (around 100 Hz)
is below the fundamental mode (the order of 10 kHz).
The cut-off frequency, $1/(2 \pi \tau_n)$, 
is extremely smaller than the 
resonant frequency, $\omega_n/(2 \pi)$, as shown in Fig. \ref{Xfreq}, 
because sound is generally faster than heat conduction.  
The absolute value $|\tilde{q}_n(\omega)|$ is 
inversely proportional to the frequency between  
$1/(2 \pi \tau_n)$ and $\omega_n/(2 \pi)$. 
Below the cut-off frequency, $1/(2 \pi \tau_n)$, 
$|\tilde{q}_n(\omega)|$ is constant. 

The frequency dependence of $\tilde{X}(\omega)=\sum \tilde{q}_n(\omega)$ 
is as follows.
The "highest" mode in Fig. \ref{Xfreq} is that with 
a wavelength comparable to the beam radius, $r_{0}$, and 
contributions of higher modes 
are negligible in the summation of Eq. (\ref{tilde X})
\cite{Gillespie,Bondu-mode}. 
The thermal relaxation time, $\tau_{r_0}$, for this "highest" mode is 
described as 
\begin{equation}
\tau_{r_0} = \frac{\rho C {r_0}^2}{4 \kappa}
\label{tau r0}
\end{equation}
from Eq. (\ref{tau n}).
The cut-off frequency of the "highest" mode, $1/(2 \pi \tau_{r_{0}})$, 
is smaller than the fundamental mode resonant frequency,
$\omega_1/(2 \pi)$, in general, as shown in Fig. \ref{Xfreq}. 
The cut-off frequency, $1/(2\pi\tau_n)$, of the lower mode is smaller 
from Eq. (\ref{tau n}).
In the range between $1/(2 \pi \tau_{r_{0}})$ and 
$\omega_1/(2 \pi)$ (the "high" frequency region in Fig. \ref{Xfreq}), 
$|\tilde{X}(\omega)|=|\sum \tilde{q}_n(\omega)|$ is inversely proportional 
to the frequency. 
Below the cut-off frequency of the fundamental mode, 
$1/(2 \pi \tau_1)$ (the "low" frequency region in Fig. \ref{Xfreq}), 
$|\tilde{X}(\omega)|$ 
is independent of the frequency. 
From Eq. (\ref{tau n}), the relaxation time, $\tau_1$, is described as 
\begin{equation}
\tau_1 = \frac{\rho C {\lambda_1}^2}{4 \kappa} = \frac{\rho C {R}^2}{\kappa},
\label{tau 1} 
\end{equation}
because
the wavelength of the fundamental mode, $\lambda_1$, is comparable 
to the mirror diameter, 2$R$.
%In the middle frequency region, 
%$1/(2 \pi \tau_1) < f < 1/(2 \pi \tau_{r_{0}})$, the general discussion 
%is difficult because the frequency dependence of the contribution of 
%the mode is different from each other. 

The typical values of $|\tilde{X}(\omega)|$ 
in the "high" and "low" frequency regions of 
Fig. \ref{Xfreq} are evaluated using 
Eq. (\ref{tilde X}). 
In the "high" frequency region, $1/(2 \pi \tau_{r_{0}}) < f < 
\omega_1/(2 \pi)$, 
it is approximated as ($|\phi_n| \ll 1$ in usual cases)
\begin{eqnarray}
|\tilde{X}(\omega)| &\sim& \frac1{2 \pi} \frac{Y\alpha}{1-2\sigma}
\frac1{\rho C}
\frac{dE}{dl}
\frac1{\omega}
\left|\sum_n \frac1{m_n {\omega_n}^2} 
\int {\rm div} \boldsymbol{w}_n dl \right|,\nonumber\\
\ \ \  &&1/(2 \pi \tau_{r_{0}}) < f < \omega_1/(2 \pi).
\label{tilde X 2}
\end{eqnarray}
The sign of the integral in Eq. (\ref{tilde X 2}) depends on the modes. 
The typical absolute value of the summation in Eq. (\ref{tilde X 2}) 
is evaluated as 
the square root of a summation of squares of the terms. 
Equation (\ref{tilde X 2}) is rewritten as 
\begin{eqnarray}
|\tilde{X}(\omega)| &\sim& \frac1{2 \pi} \frac{Y\alpha}{1-2\sigma}
\frac1{\rho C}
\frac{dE}{dl}
\frac1{\omega}\nonumber\\
&\times&\sqrt{\sum_n \frac1{{m_n}^2 {\omega_n}^4} 
\left(\int {\rm div} \boldsymbol{w}_n dl\right)^2},\nonumber\\
\ \ \ &&1/(2 \pi \tau_{r_{0}}) < f < \omega_1/(2 \pi). 
\label{tilde X 3}
\end{eqnarray}
The integral along the cosmic-ray track in Eq. (\ref{tilde X 3}) 
is evaluated as follows. 
The average of 
the length of the cosmic-ray track 
is comparable to the radius of the mirror, $R$. 
The average of $|\boldsymbol{w}_n|^2$, $\left<|\boldsymbol{w}_n|^2\right>$, 
is related to Eq. (\ref{effective mass}),
\begin{equation}
m_n = \int \rho |\boldsymbol{w}_n|^2 dV = M \left<|\boldsymbol{w}_n|^2\right>,
\end{equation}
where $M$ is the mass of the mirror. This equation gives 
\begin{equation}
\sqrt{\left<|\boldsymbol{w}_n|^2\right>} = \sqrt{\frac{m_n}{M}}.
\end{equation}
The divergence of $\boldsymbol{w}_n$ in Eq. (\ref{tilde X 3}) 
can be represented by the product of $\boldsymbol{w}_n$ and 
the wavenumber $\omega_n/v$, because $\boldsymbol{w}_n$ is the basis of the 
solution of the wave equation. 
Consequently, Eq. (\ref{tilde X 3}) is described as 
\begin{eqnarray}
|\tilde{X}(\omega)| &\sim& \frac1{2 \pi} 
\frac{Y\alpha}{1-2\sigma}\frac1{\rho C}
\frac{dE}{dl} \frac{R}{v \sqrt{M}}
\frac1{\omega}
\sqrt{\sum_n \frac1{m_n {\omega_n}^2}},\nonumber\\
\ \ \ &&1/(2 \pi \tau_{r_{0}}) < f < \omega_1/(2 \pi).
\label{tilde X 4}
\end{eqnarray}
The summation in Eq. (\ref{tilde X 4}) is the same 
as the response of a mirror against 
a static force \cite{Bondu, Bonducomment}, 
\begin{equation}
\sum_n \frac1{m_n {\omega_n}^2} = \frac{1-\sigma^2}{\sqrt{\pi}Yr_0}.
\label{Bondu}
\end{equation}
In order to simplify the discussion, the relation  
\begin{equation}
M=\pi \rho R^3
\label{M-R}
\end{equation}
is assumed. The radius of the mirror, $R$, 
is nearly equal to its thickness, $H$, 
in usual cases of interferometric gravitational 
wave detectors.
Using Eqs. (\ref{sound velocity}), (\ref{Bondu}) and (\ref{M-R}), 
Eq. (\ref{tilde X 4}) 
is written in the form 
\begin{eqnarray}
|\tilde{X}(\omega)| &\sim& \frac1{2 \pi^{7/4}}
\frac{\alpha\sqrt{1-\sigma^2}}{1-2\sigma}\frac1{\rho C}
\frac{dE}{dl} \frac1{\sqrt{R r_0}}
\frac1{\omega},\nonumber\\
\ \ \ &&1/(2 \pi \tau_{r_{0}}) < f < \omega_1/(2 \pi).
\label{tilde X approx 1}
\end{eqnarray}

Equation (\ref{tilde X}) in the "low" frequency band of 
Fig. \ref{Xfreq} is evaluated 
in the same manner as in the previous paragraph. Using Eq. (\ref{tau n}), 
the result is written as 
\begin{eqnarray}
|\tilde{X}(\omega)| &\sim& \frac1{2 \pi} \frac{Y\alpha}{1-2\sigma}
\frac1{\rho C}
\frac{dE}{dl}\nonumber\\
&&\times\sqrt{\sum_n \frac{{\tau_n}^2}{{m_n}^2 {\omega_n}^4}
\left(\int {\rm div} \boldsymbol{w}_n dl\right)^2}\nonumber\\
&\sim& \frac{\sqrt{\pi}}{2}
\frac{Y^{3/2}\alpha}{1-2\sigma}\frac1{\rho\kappa}
\frac{dE}{dl}\frac1{\sqrt{R}}
\sqrt{\sum_n \frac1{m_n {\omega_n}^6}},\nonumber\\
\ \ \ &&f < 1/(2 \pi \tau_1). 
\label{tilde X 5}
\end{eqnarray}
It can be seen that only the fundamental mode is dominant, 
because of the frequency dependence 
of ${\omega_n}^6$. The quantities of this mode 
are as follows \cite{Liu,Gillespie}: 
\begin{eqnarray}
m_1 &\sim& \frac{M}{2} \sim \frac{\pi \rho R^3}{2},
\label{mass 1}\\
\omega_1 &\sim& \frac{\pi}{H}\sqrt{\frac{Y}{\rho}}
\sim \frac{\pi}{R}\sqrt{\frac{Y}{\rho}}. 
\label{omega 1}
\end{eqnarray}
Equation (\ref{tilde X 5}) is expressed as 
\begin{equation}
|\tilde{X}(\omega)| \sim \frac1{\sqrt{2}\pi^3} 
\frac{\alpha}{1-2\sigma}\frac1{\kappa}
\frac{dE}{dl}R,
\ \ \ f < 1/(2 \pi \tau_1).
\label{tilde X approx 2}
\end{equation}

\subsection{Formula in time domain}

\begin{figure}
\begin{minipage}{8.6cm}
\includegraphics[width=8.6cm]{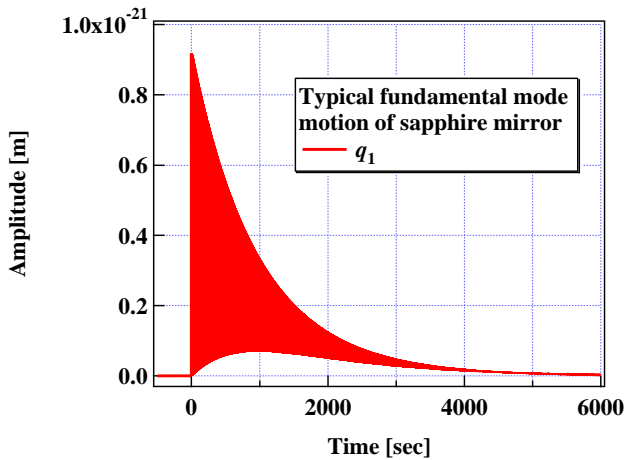}
\end{minipage} 
\caption{\label{decay}Vibration of the fundamental 
mode caused by a low-energy 
cosmic-ray muon in the time domain, $q_1$ 
in Eq. (\ref{X approx 1}). 
In the calculation, the material values of sapphire 
at room temperature are used.
Since the Q-value is extremely high, i.e. the decay time is longer 
than the period of the 
resonant motion, we are not able to see the resonant motion 
of the one period in 
this graph.}
\end{figure}

Here, we discuss the excitation formula Eq. (\ref{tilde X}) in the time domain.
Only a contribution of the $n$-th mode is considered 
in order to simplify the discussion. 
If the Q-value, $Q_n=1/\phi_n(\omega_n)$, is larger than unity 
and $1/\tau_n$ is  
smaller than $\omega_n$, $q_n$ in the time 
domain ($t>0$) is written in a form 
\begin{eqnarray}
q_n(t) &\sim& 
\frac{F_n(0)}{m_n {\omega_n}^2}
\exp\left(-\frac{t}{\tau_n}\right)\nonumber\\
&&- \frac{F_n(0)}{m_n {\omega_n}^2}
\cos(\omega_n t)\exp\left(-\frac{\omega_n t}{2Q_n}\right).
\label{X approx 1}
\end{eqnarray}
Figure \ref{decay} shows the fundamental mode, $q_1$, in the time domain. 

The second term in Eq. (\ref{X approx 1}) is dominated by 
$\tilde{X}(\omega)$ near the resonant frequency. 
This is the excited 
resonant vibration and its decay. The outputs of resonant detectors 
are described with this term. The first term in Eq. (\ref{X approx 1}) 
represents 
the drift of the center of the resonant vibration caused by the 
relaxation of thermal stress. This is dominated by $\tilde{X}(\omega)$ 
below the fundamental mode. The outputs of interferometric detectors 
are described with this term. 

The initial amplitude of $\sum_n 2F_n(0)/(m_n {\omega_n}^2)$ is 
evaluated as Eq. (\ref{tilde X approx 1}), 
\begin{eqnarray}
\sum_n \frac{2F_n(0)}{m_n {\omega_n}^2} &=& 
\frac2{\pi^{3/4}}
\frac{\alpha\sqrt{1-\sigma^2}}{1-2\sigma}\frac1{\rho C}
\frac{dE}{dl} \frac1{\sqrt{R r_0}}\nonumber\\
&=& 6.4 \times 10^{-21}\ {\rm m} 
\left(\frac{\alpha}{5 \times 10^{-6}\ {\rm /K}}\right)\nonumber\\
&&\times\left(\frac1{2.3}\frac{\sqrt{1-\sigma^2}}{1-2\sigma}\right)\nonumber\\
&&\times\left(\frac{7.9 \times 10^2\ {\rm J/kg/K}}{C}\right)\nonumber\\
&&\times\left(\frac1{2\ {\rm MeV/(g\ cm}^{-2})}\frac1{\rho}\frac{dE}{dl}\right)
\left(\frac{25\ {\rm cm}}{2R}\right)^{1/2}\nonumber\\
&&\times\left(\frac{3\ {\rm cm}}{r_0}\right)^{1/2}.
\label{typical initial amplitude}
\end{eqnarray}
%The typical value of the decay time of the excitation is considered 
%in the next section.
Here, we consider a sapphire mirror at room temperature.

\section{Discussion about the formula}

\subsection{Effect on low-temperature interferometers}
\label{low temperature discussion}

\begin{figure}
\begin{minipage}{8.6cm}
\includegraphics[width=8.6cm]{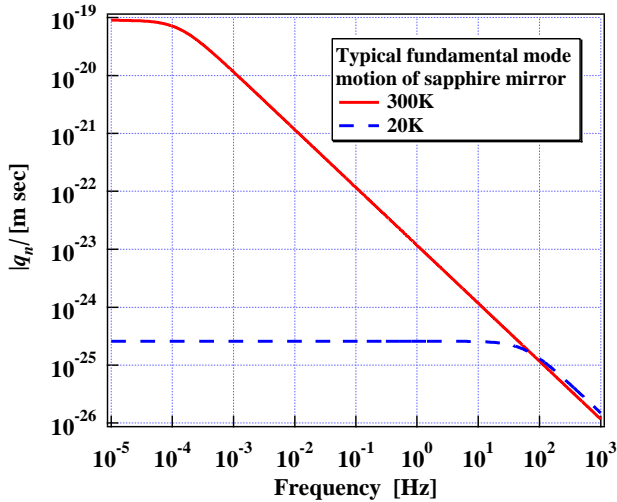}
\end{minipage} 
\caption{\label{cooling}Fourier components of 
the fundamental mode motion, $|\tilde{q}_1(\omega)|$, of a
sapphire mirror excited by a low-energy cosmic-ray muon 
at 300 K (solid line) and 20 K (dashed line). 
The cut-off frequency, $1/(2 \pi \tau_1)$, 
at 20 K is higher than that at 300 K. 
The mirror cooling reduces the low-frequency component 
because the decay time of $F_1$, which is related to 
the thermal stress, is shorter. 
The higher frequency component is independent of temperature because of the 
Gr\"{u}neisen relation \cite{Kittel}.}
\end{figure}

In some future projects using interferometric detectors as 
LCGT \cite {LCGT} and ET \cite{ET}, 
mirrors will be cooled in order to reduce the thermal noise
(for example, LCGT mirrors at 20 K). 
In the quantities of the force, $F_n$, which is related to the thermal stress, 
$\alpha,C$ and $\kappa$ in Eqs. (\ref{stress-amplitude}) and (\ref{tau n}) 
strongly 
depend on the temperature \cite{dE/dl 1}. 
The initial value of the force, $F_n(0)$ in Eq. (\ref{stress-amplitude}), 
is proportional to $\alpha/C$. The decay time of the force, 
$\tau_n$ in Eq. (\ref{tau n}), is proportional to $C/\kappa$.
The Gr\"{u}neisen relation \cite{Kittel} predicts that the ratio $\alpha/C$ 
is independent of the temperature. The initial force, $F_n(0)$, 
and the initial amplitude 
of the excited vibration do not depend on temperature.
On the contrary, in the case of crystals, 
the decay time, $\tau_n$ ($\propto C/\kappa$), 
of the first term in Eq. (\ref{X approx 1})  
is extremely short at the cryogenic temperature, 
because of the small $C$ and large $\kappa$. 
The cut-off frequency, 
$1/(2\pi \tau_n)$, in the low-temperature region is higher than 
that at room temperature (e.g. Ref. \cite{Braginsky-cosmic2}).
For example, the cut-off frequencies of sapphire at room temperature, 
obtained from Eqs. (\ref{tau r0}) and (\ref{tau 1}), are: 
\begin{eqnarray}
\frac1{2\pi \tau_1} &=& 0.13 \ {\rm mHz} 
\left(\frac{4\ {\rm g/cm}^3}{\rho}\right)
\left(\frac{7.9 \times 10^2\ {\rm J/kg/K}}{C}\right)\nonumber\\
&&\times\left(\frac{25\ {\rm cm}}{2R}\right)^2
\left(\frac{\kappa}{40\ {\rm W/m/K}}\right),\\
\frac1{2\pi \tau_{r_0}} &=& 9.0\ {\rm mHz} 
\left(\frac{4\ {\rm g/cm}^3}{\rho}\right)
\left(\frac{7.9 \times 10^2\ {\rm J/kg/K}}{C}\right)\nonumber\\
&&\times\left(\frac{3\ {\rm cm}}{r_0}\right)^2
\left(\frac{\kappa}{40\ {\rm W/m/K}}\right).
\end{eqnarray}
The values at 20 K are: 
\begin{eqnarray}
\frac1{2\pi \tau_1} &=& 58 \ {\rm Hz} 
\left(\frac{4\ {\rm g/cm}^3}{\rho}\right)
\left(\frac{0.69\ {\rm J/kg/K}}{C}\right)\nonumber\\
&&\times\left(\frac{25\ {\rm cm}}{2R}\right)^2
\left(\frac{\kappa}{1.6 \times 10^{4}\ {\rm W/m/K}}\right),
\label{tau 1 20K}\\
\frac1{2\pi \tau_{r_0}} &=& 4.0\ {\rm kHz} 
\left(\frac{4\ {\rm g/cm}^3}{\rho}\right)
\left(\frac{0.69\ {\rm J/kg/K}}{C}\right)\nonumber\\
&&\times\left(\frac{3\ {\rm cm}}{r_0}\right)^2
\left(\frac{\kappa}{1.6 \times 10^{4}\ {\rm W/m/K}}\right).
\label{tau r0 20K}
\end{eqnarray}
At low temperature, the cut-off frequencies are 
near the observation band of gravitational wave detectors (around 100 Hz). 
The "high" frequency approximation of $|\tilde{X}(\omega)|$ in 
Eq. (\ref{tilde X approx 1}) is only appropriate for room-temperature
interferometers, and not valid for cryogenic interferometers. 
In order to show the effect of the cooling mirrors, 
$|\tilde{q}_1(\omega)|$ of a
sapphire mirror at 300 K and 20 K are plotted in Fig. \ref{cooling}.
In the low-frequency region, $\tilde{X}(\omega)$ becomes much smaller due to  
cooling. The "low" frequency approximation of $|\tilde{X}(\omega)|$ in  
Eq. (\ref{tilde X approx 2}) is proportional to 
$\alpha/\kappa = \alpha/C \times C/ \kappa \propto C/\kappa$, 
which is small in 
the low-temperature region. 
Since the decay time of the force $F_n$ becomes shorter, 
it is difficult to excite the low-frequency component. 
The Fourier components $|\tilde{q}_1(\omega)|$ in the high-frequency region 
of Fig. \ref{cooling} are comparable in the cases of 20 K and 300 K.
The "high" frequency approximation of $|\tilde{X}(\omega)|$ in 
Eq. (\ref{tilde X approx 1}) is independent of temperature 
because it is proportional to $\alpha/C$ and 
related to the first term of Eq. (\ref{X approx 1}) at $t \sim 0$. 
From the cut-off frequencies in Eqs. (\ref{tau 1 20K}) and (\ref{tau r0 20K}) 
and a comparison between Eq. (\ref{tilde X approx 2}) at 20 K 
and Eq. (\ref{tilde X approx 1}), it can be seen that 
the vibration of a cooled 
sapphire mirror excited by a cosmic-ray particle 
in the observation band (around 100 Hz)
is a few-times smaller than that at room temperature. 
This is an advantage of cryogenic interferometers in addition to 
the suppression of the thermal noise 
\cite{Uchiyama2,Uchiyama3,Yamamoto-coating}, 
thermal lensing effect \cite{Tomaru} and parametric instability 
\cite{Yamamoto-PI}.  
It must be noted that motions excited by cosmic-ray particles 
in resonant detectors are independent of 
temperature \cite{Astone2,Astone3,RAP}. 
This is because the initial 
amplitude, the second term of Eq. (\ref{X approx 1}) at $t \sim 0$, 
does not depend on temperature.   

%The Gr\"{u}neisen relation \cite{Kittel} predicts that the ratio $\alpha/C$ 
%is independent of temperature. The initial force $F_n(0)$ and  
%initial amplitude of the excited vibration, 
%i.e. both the terms in Eq. (\ref{X approx 1}) at $t \sim 0$, 
%do not depend on temperature.  
%The excitation of the resonant vibration by a cosmic-ray particle, 
%the second term of Eq. (\ref{X approx 1}) at $t \sim 0$, 
%is triggered in the burst gravitational wave search with the resonators. 
%The amplitude of the motion by a cosmic-ray particle
%in the output of the burst search filter in the cryogenic resonators 
%is the comparable to that at room temperature \cite{Astone2,Astone3,RAP}.  

\subsection{Cosmic-ray track position dependence}
%\subsection{Correlations between modes}

\begin{figure}
\begin{minipage}{8.6cm}
\includegraphics[width=8.6cm]{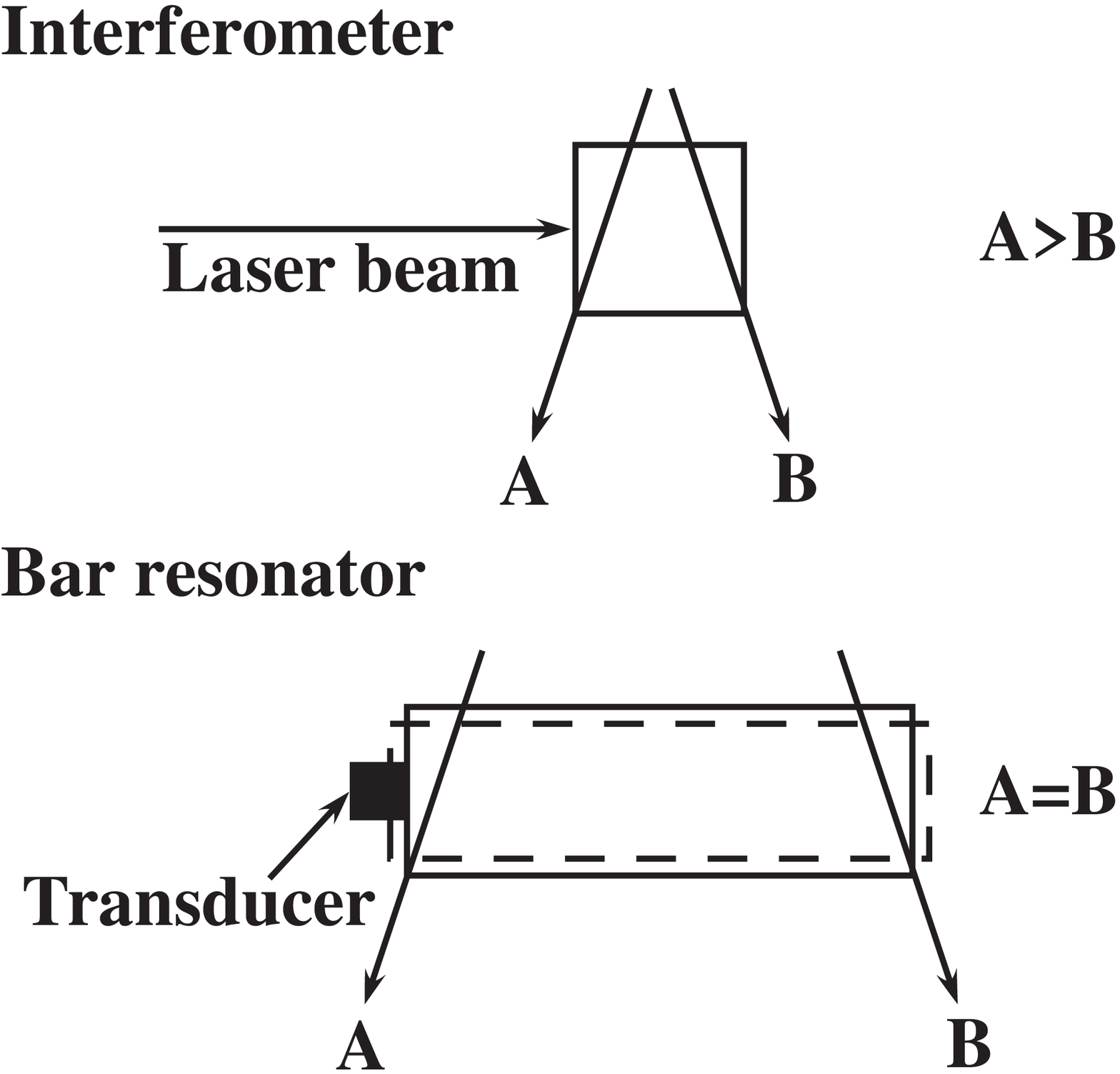}
\end{minipage} 
\caption{\label{position}Track position dependence 
of the cosmic-ray heating effect 
of interferometers and bar resonators. 
In the outputs of interferometers, the vibration caused by a particle 
that goes 
along track A near the beam spot is larger than 
that along track B far from the beam spot. On the contrary, 
in the case of bars, 
the vibration by track A 
is the same as that by track B because the displacement of 
the resonant modes is symmetric or antisymmetric 
with respect to the center of the  
bar. The dashed line shows the fundamental mode deformation.}
\end{figure}

In the calculation of Eq. (\ref{tilde X}), 
the signs of the integral terms 
in the summation are important. 
The sign depends 
on the positions of a cosmic-ray track and the laser beam spot,
because the displacement of the mode $\boldsymbol{w}_n$ 
is normalized to satisfy Eq. (\ref{normalized condition}) 
\cite{normalized condition}. 
If the particle track is near the beam spot,
the signs of the integrals of many modes are the same, because 
the basis functions, $\boldsymbol{w}_n$, on the track 
are similar.  
If the track is far from the spot, $\boldsymbol{w}_n$ on the track and 
the integral signs
are different for various modes. 
Since the sign of $q_n$ in Eq. (\ref{tilde X}) below the fundamental mode 
is the same as that of the integral, $|\tilde{X}(\omega)|$  
below the first mode is larger and smaller 
if the cosmic-ray track is near and far from the beam spot, 
respectively. Another explanation about the cosmic-ray track position 
dependence is as follows. The heated volume on the particle track pushes 
around them. 
Since the center of the mirror does not move because of the conservation 
of momentum, a larger motion is observed if the track is 
near the beam spot. 
The track position dependence of the cosmic-ray heating effect 
in interferometers 
is different from that in bar resonators, as shown in Fig. \ref{position}. 
In the outputs of bars, the vibration caused by a particle 
that goes along track A 
in Fig. \ref{position} is the same as that along track B, 
because the displacement of the 
resonant modes is symmetric or antisymmetric with respect to the center of  
the bar (the dashed line in Fig. \ref{position} shows 
the fundamental mode deformation). 
The difference between interferometers and the bar resonators is 
related to the number of modes to be considered. 
In the case of bars, only the fundamental mode 
is taken into account. On the other hand, 
in the case of interferometers, many modes contribute
to the response of a mirror. The signs of these modes have an important role. 
The discussion above is the same as that about thermal noise 
below the fundamental mode caused by 
inhomogeneously distributed loss \cite{Yamamoto-D,Yamamoto-mode,Levin}. 

\section{Application I --- Low-energy cosmic-ray particles}

%The formula of the vibration caused by a cosmic-ray particle 
%is applied to 
%the actual interferometer in the left parts of this paper. 
%\subsection{Outline of low-energy cosmic-ray particles}

\subsection{Low-energy cosmic-ray particles and interaction with matter}

%\subsection{Components and flux of cosmic-ray particles}

Primary cosmic rays generate extensive air showers 
in the atmosphere. Cosmic-ray particles on the ground are 
secondaries from air showers.
Three quarters of secondary particles at sea-level 
are muons. The remainder are almost electrons \cite{Hillas}. 
Muons with small energy (less than about 0.22 GeV) 
and electrons can be neglected because 
it is difficult to penetrate matter around the mirrors, 
for example, walls of buildings, vacuum chambers \cite{Hillas}.
The speed of muons that arrive at the mirrors 
is comparable to that of light. 
The flux of these cosmic-ray muons 
at sea-level is about $2 \times 10^{-2}\ /{\rm cm}^2/{\rm sec}$ 
\cite{Hillas,Grieder}. 

%\subsection{Interaction with matter}

Since the number of higher energy muons is smaller \cite{Gaisser}, 
the energy of most of the cosmic-ray muons is below 100 GeV. 
In this low-energy region, the dissipation process in material 
is dominated by ionization \cite{Gaisser,Longair}, 
which is Coulomb scattering with  
electrons in atoms of matter \cite{Longair}. 
The ionization loss is about 
\begin{equation}
\frac1{\rho}\frac{dE}{dl} = 2\ {\rm MeV/(g\ cm}^{-2}),
\label{energy loss}
\end{equation} 
and almost independent 
of the particle energies \cite{Gaisser, Longair}.
The typical loss per unit length, $dE/dl$, is several MeV/cm. 

%\subsection{Expected arrival interval of cosmic-ray particles
%and decay time of resonant vibration}

The effect of mirror excitation by cosmic-ray particles depends on the 
arrival frequency of particles and the decay time of the 
vibrations. If the decay time is longer than the interval of 
the particle arrivals, the mirror vibration is maintained.
If the next 
muon comes after the vibration has disappeared, 
the vibration can be treated 
as a burst event. 
The number of muons, $N$, that hit a mirror at sea-level 
per unit time is expressed as 
\begin{eqnarray}
N&=&2 \times 10^{-2}\ /{\rm cm}^2/{\rm sec} 
\times 2R \times H\nonumber\\ 
&=& 8\ {\rm /sec} \left(\frac{2R}{25\ {\rm cm}}\right)
\left(\frac{H}{15\ {\rm cm}}\right).
\label{sea-level-/sec}
\end{eqnarray}
The average arrival interval of muons, $1/N$, is 
\begin{equation}
\frac1{N}=0.13\ {\rm sec} \left(\frac{25\ {\rm cm}}{2R}\right)
\left(\frac{15\ {\rm cm}}{H}\right).
\label{sea-level-sec}
\end{equation}
The decay time of the fundamental resonant vibration is described as 
\begin{equation}
\frac{Q_1}{\pi f_1} = 8 \times 10^2\ {\rm sec} 
\left(\frac{40\ {\rm kHz}}{f_1}\right)
\left(\frac{Q_1}{10^8}\right).
\end{equation}
%The relaxation time of the thermal stress of the fundamental mode 
%is related to 
%Eq. (\ref{tau 1}):
%\begin{equation}
%\tau_1 = 3 \times 10^3\ {\rm sec}
%\left(\frac{\rho}{2\ {\rm g/cm}^3}}\right)
%\left(\frac{C}{10^3\ {\rm J/kg/K}}\right)
%\left(\frac{2R}{25\ {\rm cm}}\right)^2
%\left(\frac{10\ {\rm W/m/K}}{\kappa}\right)
%\end{equation}
Since the Q-values of mirrors used for gravitational wave detectors 
are at least $10^6$, 
the decay time is extremely larger than the expected arrival interval 
of cosmic-ray particles. 

\subsection{Power spectral density}

The power spectral density, $G_{\rm cos}(f)$, 
of vibrations caused by low-energy cosmic-ray particles has been calculated 
(e.g. Refs. \cite{Giazotto, Braginsky-cosmic2}). 
It is assumed that arrival time of particles 
and track position in a mirror are at random. 
Since there are four mirrors in an interferometer, 
the one-side power spectral density of the noise of an interferometer output 
is written in the form \cite{spectrum}
\begin{equation}
G_{\rm cos}(f) = \frac{4}{L^2} \times 8 \pi^2 N 
\left<|\tilde{X}(\omega)|^2\right> 
= \frac{32 \pi^2 N}{L^2} \left<|\tilde{X}(\omega)|^2\right>,
\label{G cosmic}
\end{equation}
where $L$ is the length of the interferometer arms. 
%The quantitiy, $N$ at sea-level, is shown in Eq. (\ref{sea-level-/sec}). 
The quantity $\left<|\tilde{X}(\omega)|^2\right>$ is the ensemble average 
of $|\tilde{X}(\omega)|^2$, which is the vibration caused by a muon. 
To evaluate the power spectrum of 
room-temperature interferometers, the square of 
Eq. (\ref{tilde X approx 1}) is used as the ensemble average, 
because this formula is appropriate to calculate the typical 
$|\tilde{X}(\omega)|$ 
at 300 K and around 100 Hz, as shown in Sec. \ref{low temperature discussion}. 
From Eqs. (\ref{energy loss}), (\ref{sea-level-/sec}) and (\ref{G cosmic}),
the power spectrum of room-temperature sapphire at sea-level is written as 
\cite{RsimH}
\begin{eqnarray}
\sqrt{G_{\rm cos}(f)} &=& 1.3 \times 10^{-26}\ /\sqrt{\rm Hz}
\left(\frac{3\ {\rm km}}{L}\right)
\left(\frac{\alpha}{5 \times 10^{-6}\ /{\rm K}}\right)\nonumber\\
&&\times\left(\frac1{2.3}\frac{\sqrt{1-\sigma^2}}{1-2\sigma}\right)
\left(\frac{7.9\times10^2\ {\rm J/kg/K}}{C}\right)\nonumber\\
&&\times\left(\frac{2R}{25\ {\rm cm}}\right)^{1/2}
\left(\frac{3\ {\rm cm}}{r_0}\right)^{1/2}
\left(\frac{100\ {\rm Hz}}{f}\right).
\label{sapphire 300K low energy}
\end{eqnarray}
The sensitivity of future second-generation interferometer projects, 
such as LCGT \cite{LCGT} and Advanced LIGO \cite{LIGO II}, is on
the order of $10^{-24}\ /\sqrt{\rm Hz}$ at 100 Hz. 
Therefore, the effect of low-energy 
cosmic-ray particles is not a serious 
problem, even in these future projects.

\section{Application II --- Shower}

High-energy cosmic-ray particles often generate many particles (showers). 
From Eqs. (\ref{energy loss}) and (\ref{SNRinterferometer2}), 
if 1000 shower particles pass in a mirror at the same time, 
the excited vibration is large enough to be detected 
by future second-generation interferometers, 
such as LCGT \cite{LCGT} and Advanced LIGO \cite{LIGO II} 
(e.g. Refs. \cite{Giazotto,Clay,Braginsky-cosmic}).  
Such excitations caused by cosmic-ray showers have been observed 
in a resonator \cite{Astone1}. 

We investigated the effect of a shower generated by a high-energy muon 
inside a mirror with a Monte-Carlo technique \cite{Monte Carlo}. 
It was assumed that the material is sapphire.  
We evaluated the probability that a high-energy muon that runs in a 30 cm 
thickness sapphire generates more than 1000 electrons. In this simulation,  
the flux of muons at sea-level 
was expressed as \cite{muon flux 1TeV}
\begin{equation}
I_{\mu}(>E)=1.1 \times 10^{-6}\ /{\rm cm}^2/{\rm sec}
\left(\frac{E}{1\ {\rm TeV}}\right)^{-2.7},
\label{mu energy spectrum}
\end{equation}
if $E$ is more than 1 TeV. Our simulation showed that the number 
per unit time and per a mirror of 
muons that generate more than 1000 electrons 
$N({\rm >1000e})$ is  
(the typical energy of such muons is about 10 TeV \cite{Monte Carlo2})
%\begin{equation}
%I_{\mu}({\rm >1000e}) 
%= 2.3 \times 10^{-14}\ /{\rm cm}^2/{\rm sec}.
%\label{1000electorn intensity}
%\end{equation}
\begin{equation}
N({\rm >1000e}) 
%= I_{\mu}({\rm > 1000e}) \times 2R \times H 
= 1.0 \times 10^{-11}\ {\rm /sec} \left(\frac{2R}{25\ {\rm cm}}\right)
\left(\frac{H}{15\ {\rm cm}}\right).
\label{1000TeV-/sec}
\end{equation}
It must be noted that this average arrival number, $N({\rm >1000{\rm e}})$, 
was overestimated, because 
only a part of muons has more than a 
30 cm length track in a sapphire mirror. 
Since there are four mirrors in an interferometer, 
the average arrival interval is 
\begin{equation}
\frac1{4 N({\rm >1000 e})}=7.8 \times 10^2 \ {\rm year} 
\left(\frac{25\ {\rm cm}}{2R}\right)
\left(\frac{15\ {\rm cm}}{H}\right).
\label{1000TeV-sec}
\end{equation}
The effect of showers generated by high-energy muons inside mirrors 
is not a serious problem.

In the case of a shower that occurs near a mirror, 
the energy of an original particle that generates 1000 particles
is about 1 TeV \cite{Braginsky-cosmic}. 
Since the spread of particles in a TeV energy shower is quite large,
the typical size mirror of interferometers can not contain all of  
the energy of a thousand particles.
In order to know how often more than 1000 particles go into 
a mirror, accurate simulations about shower generation 
in apparatus around mirrors (for example, vacuum chambers and vibration
isolation systems) and 
the response of a mirror are necessary as resonators 
\cite{Ricci,Chiang,Coccia}. This is our future work.   

\section{Application III --- Exotic-particle search}

The effect of cosmic-ray particles on gravitational wave detectors 
suggests that the detectors are useful to search for exotic particles 
that dissipate a large amount of energy in material. 
Ideas that resonators can be used 
as magnetic monopole \cite{Bernard} or 
mirror dust particle \cite{Foot}
detectors were proposed. The upper limits of the flux of 
nuclearite \cite{Witten,Rujula} from the operation of 
resonators were reported \cite{Liu, Astonenuclearite}. 
Here, we discuss interferometers as exotic-particle detectors 
in comparison with resonators 
(bars \cite{ALLEGRO,EXPLORER,NAUTILUS,AURIGA,NIOBE}). 

%\subsection{Comparison between interferometer and bar resonator}

In order to detect exotic particles or other rare events,
a larger aperture and 
higher sensitivity are required for detectors. 
The cross section of a bar resonator is 10-times larger than 
that of an interferometer \cite{cross section}. 
The area of a bar is about 1.8 m$^2$ 
(diameter, 0.6 m; length, 3 m) \cite{Astonenuclearite}. 
The cross section of four mirrors of an interferometer 
is about 0.15 m$^2$ (diameter, 0.25 m; thickness, 0.15 m). 
We discuss the sensitivity of interferometers and bars for an 
exotic particle passage.

\subsection{Signal-to-Noise ratio of interferometers} 
%\subsection{Definition of S/N} 

Since the time evolution of an excited motion by an exotic particle 
is predicted from Eq. (\ref{tilde X}), 
the matched filtering method can be applied to the outputs of detectors. 
The output of a matched filter, the signal-to-noise ratio (S/N), 
is defined as \cite{SNR}
\begin{equation}
{\rm S/N} = 4 \pi 
\sqrt{\int_0^{\infty} 
\frac{|\tilde{S}(\omega)|^2}{G_{\rm det}(f)}df},
\label{SNR}
\end{equation}
where $\tilde{S}(\omega)$ and $G_{\rm det}(f)$ 
are the Fourier components of the 
signal and the one-side power spectral density of the noise 
of gravitational wave detectors, 
respectively. 

In the case of interferometers, $G_{\rm det}(f)$ and $\tilde{S}(\omega)$ 
of Eq. (\ref{SNR}) are the strain noise, $G_{\rm int}(f)$, 
and the ratio of the Fourier component of the motion excited 
by an exotic particle, $\tilde{X}(\omega)$, to the arm length, $L$, 
respectively.  
It is supposed that the temperature is 300 K. 
Here, we recall Eq. (\ref{tilde X approx 1}) in the form 
\begin{eqnarray}
|\tilde{X}(\omega)| &\sim& \frac{A}{\pi^{3/4}} 
\frac{\sqrt{1-\sigma^2}}{Y\sqrt{Rr_0}}\frac1{f},
\label{tilde X approx 4}\\
A &=& \frac1{4 \pi^2}
\frac{Y\alpha}{1-2\sigma}\frac1{\rho C}
\frac{dE}{dl}.
\label{A}
\end{eqnarray}
S/N is expressed by using Eqs. (\ref{SNR}) and 
(\ref{tilde X approx 4}):  
\begin{equation}
{\rm S/N}_{\rm int} = 
\frac{4\pi^{1/4} A \sqrt{1-\sigma^2}}{Y\sqrt{Rr_0}L}
\sqrt{\int \frac{df}{f^2 G_{\rm int}(f)}}.
\label{SNRinterferometer}
\end{equation}

\subsection{S/N of bar resonators}

In the case of bar detectors, $\tilde{S}(\omega)$ 
and $G_{\rm det}(f)$ of Eq. (\ref{SNR}) 
are the force applied by an exotic particle 
to the $n$-th mode, $\tilde{F}_n(\omega)$, 
in Eq. (\ref{tilde F_n}), and the tidal force, which corresponds to 
the strain noise, $G_{\rm bar}(f)$, respectively.  
Equation (\ref{tilde F_n}) is rewritten as 
\begin{equation}
|\tilde{S}(\omega)|=|\tilde{F_n}(\omega)| = A \frac1{f} 
\left(\int {\rm div} \boldsymbol{w}_n dl \right),
\label{resonator particle force n}
\end{equation}
because the cut-off frequency, $1/(2\pi \tau_n)$, is lower than the 
resonant frequency. 
Here, we take the fundamental mode of bars into account.
Under the same approximation in the derivation of 
Eqs. (\ref{tilde X approx 1}) and (\ref{tilde X approx 2}), 
$\tilde{S}(\omega)$ 
is expressed as 
\begin{equation}
|\tilde{S}(\omega)|=|\tilde{F_1}(\omega)| 
\sim A \frac{\pi}{\sqrt{2}} \frac1{f}.
\label{resonator particle force}
\end{equation}
The tidal force that corresponds to the strain noise, $G_{\rm bar}(f)$, 
is obtained from Refs. \cite{Liu,Hirakawa,tidal},
\begin{equation}
G_{\rm det}(f)=\left(M_{\rm b} {\omega_{1({\rm b})}}^2 
\frac{l}{\pi^2} \right)^2
G_{\rm bar}(f),
\label{resonator force noise}
\end{equation}
where $M_{\rm b}$, $\omega_{1(\rm b)}$ and $l$ are the mass, 
angular resonant frequency 
of the fundamental mode and length of a bar. 
S/N is given by Eqs. (\ref{SNR}), (\ref{resonator particle force}), 
and (\ref{resonator force noise}): 
\begin{equation}
{\rm S/N}_{\rm bar} = \frac{2 \sqrt{2} \pi^{4} A}
{M_{\rm b} {\omega_{1({\rm b})}}^2 l}
\sqrt{\int \frac{df}{f^2 G_{\rm bar}(f)}}.
\label{SNRresonator}
\end{equation}

\subsection{Comparison between interferometers and bar resonators}

Here, we discuss the effects of an exotic particle on interferometers 
and bar detectors by using 
Eqs. (\ref{SNRinterferometer}) and (\ref{SNRresonator}). 
The integral term only depends 
on the sensitivity of gravitational wave detectors. 
It must be noted that the weight, $1/f^2$, originates 
from the frequency dependence of $\tilde{S}(\omega)$, 
i.e. Eqs. (\ref{tilde X approx 4}) and (\ref{resonator particle force n}). 
The integral term in Eq. (\ref{SNRinterferometer}) 
for future second-generation interferometers, 
e.g. the LCGT project \cite{LCGT}, is 
\begin{equation}
\sqrt{\int \frac{df}{f^2 G_{\rm int}(f)}} = 3.0 \times 10^{22}.
\label{integral interferometer}
\end{equation}
The typical goal sensitivity of bar resonators \cite {Vinante} 
is $\sqrt{G_{\rm bar}(f)} \sim 3\times 10^{-22}\ /\sqrt{\rm Hz}$ 
in the frequency range
between 850 Hz and 950 Hz.
% \cite{noise temperature}. 
The integral term in Eq. (\ref{SNRresonator}) is 
\begin{equation}
\sqrt{\int \frac{df}{f^2 G_{\rm bar}(f)}} \sim 3.7 \times 10^{19}.
\label{integral resonator}
\end{equation}
The integral term of interferometers is 1000-times larger, 
because interferometers have higher sensitivity 
and a wider observation band. Since the observation band of interferometers 
is lower than 
that of resonators, the weighting function, $1/f^2$, increases the 
integral term of interferometers.

The factors, except for the integral term and $A$ 
in Eqs. (\ref{SNRinterferometer}) 
and (\ref{SNRresonator}), 
are the ratios of the responses to an exotic particle to that 
to the gravitational wave. If this factor is large, 
the detector is more suitable for 
exotic-particle searches. 
This factor of interferometers is \cite{beam radius}
\begin{eqnarray}
\frac{4\pi^{1/4} \sqrt{1-\sigma^2}}{Y\sqrt{Rr_0}L} 
&=& 4.9 \times 10^{-14}\ /{\rm N}
\left(\frac{\sqrt{1-\sigma^2}}{0.96}\right)\nonumber\\
&&\times\left(\frac{4\times10^{11}\ {\rm Pa}}{Y}\right)
\left(\frac{25\ {\rm cm}}{2R}\right)^{1/2}\nonumber\\
&&\times\left(\frac{6\ {\rm cm}}{r_0}\right)^{1/2}
\left(\frac{3\ {\rm km}}{L}\right).
\label{other factor interferometer}
\end{eqnarray}
In the case of bar resonators, this factor is 
\begin{eqnarray}
\frac{2 \sqrt{2} \pi^{4}}{M_{\rm b} {\omega_{1(\rm b)}}^2 l} 
&=& 1.2 \times 10^{-9}\ /{\rm N}
\left(\frac{2300\ {\rm kg}}{M_{\rm b}}\right)\nonumber\\
&&\times\left(\frac{900 \times 2 \pi\ {\rm rad/Hz}}{\omega_{1(\rm b)}}\right)^2
\left(\frac{3\ {\rm m}}{l}\right).
\label{other factor resonator}
\end{eqnarray}
The factor of bars is extremely larger. 
The main reason for this difference comes from the sizes of the detectors, 
$L$ and $l$. 
An exotic-particle detector must be a good displacement sensor. 
A smaller size detector is a better displacement sensor, 
if the strain (gravitational wave) sensitivity is the same. 
The better strain sensitivity of interferometers shown by 
Eqs. (\ref{integral interferometer}) and (\ref{integral resonator})
is canceled by their larger size. 
The factors in Eqs. (\ref{other factor interferometer}) 
and (\ref{other factor resonator}), 
except for $L$ and $l$, represent the mechanical responses of a
mirror and a bar. The response of a bar is typically about 10-times larger.

The amplitude of the force 
%related to the thermal stress applied to the $n$-th mode 
$F_n$ in Eq. (\ref{resonator particle force n}) is proportional to $A$.
This quantity depends on only the energy loss process 
of exotic particles and the material of the mirrors and the bar resonators. 
This is evaluated as 
\begin{eqnarray}
A &=& 7.3 \times 10^{-9}\ {\rm N} 
\left(\frac{Y}{4\times10^{11}\ {\rm Pa}}\right)
\left(\frac{\alpha}{5 \times 10^{-6}\ {\rm /K}}\right)\nonumber\\
&&\times\left(\frac{0.42}{1-2\sigma}\right) 
\left(\frac{7.9\times10^2\ {\rm J/kg/K}}{C}\right)\nonumber\\
&&\times\left(\frac1{3\ {\rm GeV/(g\ cm}^{-2})}
\frac1{\rho}\frac{dE}{dl}\right).
\label{A quantitiy}
\end{eqnarray} 
In the quantities of Eq. (\ref{A quantitiy}), 
only the linear thermal-expansion coefficient,  
$\alpha$, strongly depends on the  
material \cite{monopole comment}. The values of the 
coefficient $\alpha$ for
fused silica and sapphire at 300 K, 
are $5.5 \times 10^{-7}\ /{\rm K}$ and 
$5.0 \times 10^{-6}\ /{\rm K}$, respectively. 
The coefficient $\alpha$ of the alloy Al5056 \cite{Suzuki}, 
which is the most popular material of bar resonators
\cite{ALLEGRO, EXPLORER, NAUTILUS, AURIGA}, 
is $2.3 \times 10^{-5}\ /{\rm K}$. 
%The quantitiy, $A$, of the bar 
%is 5 or 50 times larger 
%than that of the interferometer 
%owing to the large coefficient of thermal expansion.  

From the above discussion, the advantages of interferometers, 
the higher strain 
sensitivity and wider observation band, are canceled 
by their larger detector size, 
because exotic-particle detectors must have 
good displacement sensitivity, 
not strain sensitivity. 
The larger mechanical response (about 10 times) 
and linear thermal-expansion coefficient
(several or several tens times) of bar resonators enhance the sensitivity. 
The typical S/N of interferometers is obtained 
from Eqs. (\ref{SNRinterferometer}), (\ref{integral interferometer}), 
(\ref{other factor interferometer}), and (\ref{A quantitiy}):
\begin{eqnarray}
{\rm S/N}_{\rm int} &=& 10^1\ 
\left(\frac{\alpha}{5.0 \times 10^{-6}\ /{\rm K}}\right)\nonumber\\
&&\times\left(\frac1{3\ {\rm GeV/(g\ cm}^{-2})}
\frac1{\rho}\frac{dE}{dl}\right).
\label{SNRinterferometer2}
\end{eqnarray}
The S/N of bar resonators is evaluated from 
Eqs. (\ref{SNRresonator}), (\ref{integral resonator}), 
(\ref{other factor resonator}), and (\ref{A quantitiy}):
\begin{eqnarray}
{\rm S/N}_{\rm bar} &=& 3 \times 10^2\ 
\left(\frac{\alpha}{2.3 \times 10^{-5}\ /{\rm K}}\right)\nonumber\\
&&\times\left(\frac1{3\ {\rm GeV/(g\ cm}^{-2})}
\frac1{\rho}\frac{dE}{dl}\right).
\label{SNRresonator2}
\end{eqnarray}
The sensitivity for an exotic particle of bars 
is a few tens or a few hundreds times better 
than that of interferometers \cite{sensitivity}. 
The sensitivity of bars in the above discussion, 
Eq. (\ref{integral resonator}), is based on the goal sensitivity. 
The current sensitivity is 10-times worse than it \cite{Vinante}. 
The current bar resonators are the better exotic-particle 
detectors than  
the future second-generation interferometers 
as LCGT \cite{LCGT} and Advanced LIGO \cite{LIGO II}.

%\subsection{Interferometer improvement for exotic-particle search}

It is difficult to improve the sensitivity of interferometers 
for exotic particles. One reason is that  
in order to enhance the signal, 
the mechanical response and the coefficient of thermal expansion 
of a mirror must be larger. 
Equation (\ref{other factor interferometer}) implies 
that a smaller mirror and beam yield a larger mechanical response. 
However, this strategy enhances the amplitude of the displacement noise, 
and the S/N does not increase. 
A smaller mirror increases the radiation-pressure noise ($\propto R^{-3}$). 
A smaller beam and a larger coefficient of thermal expansion increase 
the amplitude of the thermal noise caused by thermoelastic damping 
in the mirror substrate ($\propto \alpha/{r_0}^{3/2}$) \cite{Braginsky}. 
Although mirror cooling reduces 
the thermal noise \cite{Uchiyama2,Uchiyama3,Yamamoto-coating}, 
S/N does not become larger,
because the excitation by an exotic particle 
becomes smaller than that at room temperature, as shown 
in Sec. \ref{low temperature discussion}.

\section{Conclusions}

%The vibration of the resonant gravitational wave detector caused 
%by the heat energy due to cosmic-ray particles was studied well. 
%On the contrary, the effect on the interferometric detector 
%was seldom investigated. In this paper, 
We obtained a general formula for a mirror vibration caused by 
a cosmic-ray particle, and studied the effects in typical cases
of interferometric experiments. 
This formula reveals differences in the responses of resonators and 
interferometers against cosmic-ray particles.  
In the case of resonators, the contribution of the resonant vibration 
is dominant. On the contrary, in the case of interferometers, 
the motion of the centers of resonant vibrations must be taken 
into account. 
Although the effect of cosmic-ray particles 
of resonators is independent of the temperature, 
in the case of interferometers, 
vibrations caused by cosmic-ray particles can be reduced by using 
cooling mirrors.  
In the case of bar resonators, the particle track position dependence 
of the vibration 
by a cosmic-ray particle is symmetric 
with respect to the center of a resonator, as shown in Fig. \ref{position}. 
On the other hand, in interferometers, larger motion is observed 
if the track is near  
the laser beam spot on the surface of a mirror. 

The typical vibration amplitude of interferometers caused 
by cosmic-ray particles was evaluated. 
The power spectrum of vibrations by low-energy 
cosmic-ray muons (less than 100 GeV) is about 100-times smaller 
than the goal sensitivity 
of the future second-generation projects, such as LCGT and Advanced LIGO. 
The arrival frequency of high-energy cosmic-ray muons 
that generate enough large showers inside the mirrors of LCGT and Advanced LIGO
is one per a millennium. 
If a shower that occurs near a mirror brings more than 
a thousand particles to the mirror 
(an original particle of the shower has an energy that is more than 1 TeV), 
the vibration will be observed in LCGT and Advanced LIGO interferometers. 
A detailed study on such shower events is our future work.
We also discussed the possibility of a use of gravitational wave detectors 
for exotic-particle searches.
Interferometers and bar resonators were compared as detectors for  
such an exotic-particle search. 
The cross section of bars 
is 10-times larger than that of interferometers. 
The sensitivity of bars for an exotic particle
is $(30 \sim 300)$ times better 
than that of interferometers. 

\begin{acknowledgments}
We are grateful to Jun Nishimura for useful comments.
\end{acknowledgments}

%\bibliography{apssamp}% Produces the bibliography via BibTeX.

\begin{thebibliography}{99}
\bibitem{Liucomment}Our result concerning the force $F_n(0)$ agrees with that 
derived from $\delta E_n$, 
which is equal to ${F_n}^2/(2 m_n {\omega_n}^2)$ 
in Eq. (2) of Ref. \cite{Liu}. 
The quantity $C_{\rm p}$ in Eq. (2) of Ref. \cite{Liu} 
is $\rho C$ in this paper. 
It must be noted that $M$ in Eq. (2) of Ref. \cite{Liu} 
is the mass of a mirror,
not the effective mass, $m_n$. This discrepancy is canceled 
by the difference of the normalized condition of the basis functions, 
$\boldsymbol{w}_n$. In this paper, the normalized condition is 
Eq. (\ref{effective mass}). In Ref. \cite{Liu}, 
the normalized condition is the same as Eq. (5.6) of Ref. \cite{Bernard}.
\bibitem{Bonducomment}This formula is valid if the mirror thickness is 
comparable to the mirror radius, and the beam radius 
is extremely smaller than the mirror radius.
\bibitem{ET}
Einstein Telescope (ET) is a future third-generation project in Europe. 
The interferometer has vacuum tubes, of which the total length is 30 km,
and cryogenic mirrors. 
The details of the design are presently being discussed  
(H. L\"{u}ck, private communications).
\bibitem{dE/dl 1}The super-conductive transition changes, $dE/dl$ 
\cite{Astone2,Astone3,RAP}. 
Such a case is not considered because the mirror material 
does not generally become a superconductor.
\bibitem{normalized condition}
The results of the 
discussion presented in this section are valid under 
arbitrary normalization conditions \cite{Yamamoto-mode}. 
The right-hand sides of 
Eqs. (8) and (10) in a general case are 
$\sum_{n} q_n(t) 
\int w_{n,{\rm opt}}(\boldsymbol{r}) 
P(\boldsymbol{r}) dS$
and $\tilde{F}_n 
/ \int w_{n,{\rm opt}}(\boldsymbol{r}) 
P(\boldsymbol{r}) dS$, respectively.
The signs of $q_n(t) \int w_{n,{\rm opt}}(\boldsymbol{r}) 
P(\boldsymbol{r}) dS$
and $\tilde{F}_n / \int w_{n,{\rm opt}}(\boldsymbol{r}) 
P(\boldsymbol{r}) dS$ cancel out each other in the process of the 
calculation for $\tilde{X}$.
\bibitem{spectrum}In this paper, the one-side power spectral density 
$G_{Y}(f)$ is defined as 
$\lim_{T \rightarrow \infty} 8\pi^2 
\left<|\int_{-T/2}^{T/2}Y(t)\exp(-{\rm i}\omega t)dt/(2\pi)|^2\right>/T$.
The integral $\int_0^{\infty} G_{Y}(f) df$ is the average of $|Y(t)|^2$,  
$\lim_{T \rightarrow \infty} \int_{-T/2}^{T/2} \left<|Y(t)|^2\right> dt/T$ 
(Parseval's  theorem).
\bibitem{RsimH}It must be remembered that Eq. (\ref{tilde X approx 1}) 
is an approximation formula 
under the condition that the mirror radius, $R$, 
is comparable to its height, $H$.
\bibitem{Monte Carlo}
The details of the Monte-Carlo simulation are as follows. 
The energy of muons is dissipated via three processes: 
bremsstrahlung 
(photon emission from 
charged particles accelerated by the Coulomb potential of nuclei), 
direct electron-pair production and interactions with nuclei \cite{Gaisser}.
Photons and electron-positron pairs originating from these processes 
generate electromagnetic cascades. 
In order to simplify the simulation, 
secondary particles generated in the interactions 
between muons and nuclei 
were replaced by a gamma-ray photon.
Since the interactions with nuclei are rather rare compared to the other 
two processes mentioned above, this simplification did not have a 
serious effect on the results. 
In the process of the electromagnetic cascade, the bremsstrahlung 
and ionization of 
electrons and the electron-positron creation by photons 
were taken into account.  
Our simulation about an electromagnetic cascade from an electron agreed with 
the result shown in Ref. \cite{Rossi}.
Moreover, the result of this Monte-Carlo simulation was consistent 
with the formula 
of the averaged energy dissipation of a high-energy muon 
(more than about 1 TeV) 
in sapphire \cite{initial2,Gaisser}, 
$dE/dl = 8\ {\rm MeV/cm}\ [E/(700\ {\rm GeV})]$.
It was assumed that sapphire consists of a kind of atom. It had 
$Z^2=10.3^2$ (square of atomic number) and $A=21$ (mass number) 
which are the averages of those of Al$_2$ and O$_3$.
\bibitem{muon flux 1TeV}In order to evaluate this expression, we used a 
standard muon intensity at sea-level given by Gaisser \cite{Gaisser} 
(equation (3) and the original Gaisser parameterization in Ref. \cite
{Tang}). 
The zenith angle dependence of this muon intensity 
was evaluated taking into account the 
radius of the Earth (6367.5 km) and the thickness of the atmosphere (32 km) 
as Fig. 12 of Ref. \cite{Muraki}. In order to check our calculation, 
we considered another formula about the zenith angle distribution, 
Eq. (10) in Ref. \cite{Okada}. We obtained the almost same result.
\bibitem{Monte Carlo2}
From our simulation, a 10 TeV muon does not usually generate 1000 electrons 
after it runs in the 30 cm thickness sapphire.  
In the average case, the electron number increases until 
the muon track length in the sapphire reaches 150 cm. 
In the stationary state after a 150 cm run, 
the average electron number is about 10 for the case of a 10 TeV muon (this 
average electron number in the stationary state is proportional 
to the muon energy).
\bibitem{cross section}
There are exotic-particle detectors that have larger cross sections 
than those of bar resonators. For example, the total cross section of 
the MACRO detector is about 1000 m$^2$ \cite{MACRO1,MACRO2}.
\bibitem{tidal}Since the result depends on the Poisson ratio weakly, 
it is supposed that this quantity is zero.
\bibitem{beam radius}In order to reduce the thermal noise 
below the goal sensitivity of the future second-generation interferometers 
without mirror cooling, 
the beam radius must be larger than the typical value 3 cm.
In Advanced LIGO \cite{LIGO II}, the beam radius is 6 cm.
\bibitem{monopole comment}The loss of monopoles in conductors 
is different from that in insulators, 
because the former is dominated by the eddy-current dissipation 
\cite{Bernard}. 
Such a case is not considered 
because the mirror material of the interferometric 
gravitational wave detectors is not usually a conductor.
\bibitem{sensitivity}
The scintillators and track-etch detectors 
(e.g. CR39) \cite{MACRO1,MACRO2,Astonenuclearite,Nakamura} 
have a sensitivity that is comparable to that of bar resonators.
From Eq. (\ref{SNRresonator2}) and Ref. \cite{Rujula,MACRO1}, 
these apparatus find exotic particles, even if the energy loss $(1/\rho)dE/dl$
is on the order of 10 MeV/(g cm$^{-2}$).
\end{thebibliography}

\end{document}